\documentclass[review]{elsarticle}

\usepackage{lineno,hyperref}
\modulolinenumbers[5]

\journal{International journal of forcasting}

\usepackage{mathtools}
\usepackage{amsfonts}

\begin{document}

\begin{frontmatter}

\title{Stock Price Prediction using Principle Components}

\author[mymainaddress]{Mahsa Ghorbani\corref{mycorrespondingauthor}}
\cortext[mycorrespondingauthor]{Corresponding author}
\ead{mahsa.ghorbani@colostate.edu}

\author[mysecondaryaddress]{Edwin K. P. Chong}

\address[mymainaddress]{Electrical and Computer Engineering department, Colorado State University}
\address[mysecondaryaddress]{Systems Engineering departmetn, Colorado State University}

\fntext[myfootnote]{A preliminary version of parts of this paper was presented in \cite{ghorbaninew}}

\begin{abstract}
The literature provides strong evidence that stock prices can be predicted from past price data. Principal component analysis (PCA) is a widely used mathematical technique for dimensionality reduction and analysis of data by identifying a small number of principal components to explain the variation found in a data set. In this paper, we describe a general method for stock price prediction using covariance information, in terms of a dimension reduction operation based on principle component analysis. Projecting the noisy observation onto a principle subspace leads to a well-conditioned problem. We illustrate our method on daily stock price values for five companies in different industries. We investigate the results based on mean squared error and directional change statistic of prediction, as measures of performance, and volatility of prediction as a measure of risk.\tnoteref{mytitlenote}
\end{abstract}


\end{frontmatter}


\section{Introduction}

Stock price prediction is one of the most challenging problems in finance. Successful forecasting of stock prices can yield significant profit in trading. Technical traders base their analysis on the promise that the patterns in market prices are assumed to recur in the future, and thus these patterns can be used for predictive purposes\cite{gencay1998predictability}. Studies show fundamental variables such as earnings yield, cash flow yield, size and book to market ratio \cite{chen2003application,fama1992cross}, and macroeconomic factors such as interest rates, expected inflation, and dividend have some power to predict stock returns\cite{ fama1988permanent, fama1988dividend}.

The literature provides strong evidence that price can also be predicted from past price/return data as well as other fundamental or macroeconomic variables. 
Some studied have found significant auto-correlation for returns over a short period of time. French and Roll find negative correlation for individual securities for daily returns \cite{french1986stock}. Others find significant positive serial correlation for weekly and monthly holding-period returns \cite{lo1988stock}. 
Studies also demonstrate stock return correlation over the horizon of several months or years. Fama and French report that the auto-correlation is weak for the daily and weekly holding periods common in market efficiency tests but stronger for long-horizon returns \cite{fama1988permanent}. They find negative serial correlation in returns of individual stocks and various portfolios over three to five year intervals. Cutler et al. show that returns are positively correlated over a horizon of several months and negatively correlated at the three to five year horizon \cite{cutler1991speculative}.
There are some other studies that also show correlation in stock returns over a multiple year interval \cite{chopra1992performance,bondt1985does} which all confirm that price/return values are predictable from past price/return values. 

In this paper, we describe a general method for predicting future stock price values based on historical price data, using covariance information. The covariance matrix of a data set is known to be well approximated with the classical maximum likelihood estimator \cite{flury1988common} or empirical covariance if the number of observations is large enough compared to the number of predictors. The problem occurs when the matrix dimension ($N$) is large relative to the number of observations ($K$). In this case the above estimators are not reliable. When $K$ is smaller than $M$, the problem is even worse because the matrix is not positive definite \cite{stein1975estimation}. This problem, which happens quite often in finance, gives rise to a new class of estimators such as shrinkage estimators. For example Ledoit and Wolf, shrink the sample covariance towards a scaled identity matrix and propose a shrinkage coefficient distribution that minimizes the Mean Squared Error between the estimated and the real covariance matrix \cite{ledoit2004well}. Some other studies in this field include  \cite{ledoit2003improved,daniels1999nonconjugate,haff1980empirical}.
In our numerical evaluations in this paper we have sufficient empirical data to produce a reliable empirical covariance matrix, at least adequate to illustrate the efficacy of our method.

Mean squared error (MSE) is considered an appropriate metric to measure the performance of predictive tools \cite{gencay1998predictability, fama1988dividend}, which is the average squared difference between the actual and predicted price value. Given historical data, one efficient way to estimate future prices is by simply using the multivariate conditional mean as the point estimator, because it minimizes the mean squared error of the estimate \cite{scharf1991statistical}. However, numerical results using this method cannot always be trusted because of associated ill-conditioning issues. Our main goal is to propose a method with similar forecasting power that does not suffer from this issue.

Principal component analysis (PCA) is a well-established mathematical procedure for dimensionality reduction and has wide applications across various fields such as time-series prediction \cite{hotelling1933analysis}, pattern recognition, feature extraction, data compression, and visualization \cite{jolliffe2002principal}. In the field of quantitative finance, PCA has relevance in exploring financial time series \cite{ince2007kernel}, dynamic trading strategies \cite{fung1997empirical}, financial risk computations \cite{alexander2009market, fung1997empirical}, and statistical arbitrage \cite{shukla1990sequential}. In this work, we employ PCA in forecasting stock prices. 

Our method bears similarity with subspace filtering methods. Such methods are based on the orthogonal decomposition of the noisy data space onto a signal subspace and a noise subspace. This decomposition is possible under the assumption of a low-rank model for the data, and on the availability of an estimate of the noise covariance matrix \cite{hermus2006review}. This task can be done based on a modified singular value decomposition (SVD) of data matrices \cite{tufts1982data}. The orthogonal decomposition into frame-dependent signal and noise subspace can be performed by an SVD of the noisy signal observation matrix or equivalently by an eigenvalue decomposition of the noisy signal covariance matrix \cite{hermus2006review}. 

We compare the performance of our proposed methods in terms of MSE and directional change statistic. Directional change statistic calculates the correct matching number of the actual and predicted values with respect to the directional change \cite{yao2000case}. It is an important evaluation measure of the performance because predicting the direction of price movement is very important in some market strategies. 

Another important parameter that we are interested in, is standard deviation, one of the key fundamental risk measures in portfolio management \cite{torun2011portfolio}. The standard deviation is a statistical measure of volatility, often used by investors to measure the risk of a stock or portfolio.

As mentioned above, in this paper we focus on forecasting stock prices from daily historical price data. In Section~\ref{part1}, we introduce our technical metholody, and in particular estimation techniques using covariance information. In Section~\ref{part2}, we describe our method for processing the data and estimating the covariance matrix from empirical data, including data normalization. We also demonstrate the performance of our method.

\section{Theoretical Methodology}\label{part1}

\subsection{Estimation Techniques}
In this section we introduce a new computationally appealing method for estimating future stock price values using covariance information. The empirical covariance can be used as an estimate of the covariance matrix if enough empirical data is available, or we can use techniques similar to the ones introduced in the previous section.

Suppose that we are given the stock price values for $M$ days. The overall goal is to predict the next $M+1$ to $N$ data points, which represents company stock prices over the next $N-M$ trading days using the observed values of the past consecutive $M$ days. The reason for introducing $N$ will be clear below.

\subsubsection{Gauss-Bayes or Conditional Estimation of $z$ given $y$}
Suppose that $x$ is a random vector of length $N$. Let $M\leq N$ and suppose that the first $M$ data points of vector $x$ represent the end-of-day prices of a company stock over the past $M$ consecutive trading days. The multivariate random vector $x$ and can be partitioned in the form

\begin{equation}
x=\begin{bmatrix}
y & z
\end{bmatrix},
\end{equation}

Let random vector $y$ representing the first $M$ data points and $z$ the price of the next $N-M$ days in the future. We wish to estimate $z$ from $y$.

The covariance matrix for the random vector $x$ can be written as
\begin{equation}\label{sigmas}
\Sigma _{xx}=\begin{bmatrix}
 \Sigma _{yy}& \Sigma _{yz}\\ 
 \Sigma _{zy}& \Sigma _{zz}
\end{bmatrix},
\end{equation}
where $\Sigma _{yy}$ is the covariance of $y$ and $\Sigma _{zz}$ is the covariance of $z$. 
Assuming that $y$ and $z$ are jointly normally distributed, knowing the prior distribution of $x=[y,z]$, the Bayesian posterior distribution of $z$ given $y$ is given by
\begin{align}\label{cond}
 \widehat {z}_{z\mid y}&=\Sigma _{zy}\Sigma _{yy}^{-1}y \nonumber\\
  \widehat \Sigma _{z\mid y}&=\Sigma _{zz}-\Sigma _{zy}\Sigma _{yy}^{-1}\Sigma _{yz}.
\end{align}

The $\widehat\Sigma_{{z\mid y}}$ matrix representing the conditional covariance of $z$ given $y$, is also called the Schur complement of $\Sigma_{yy}$ in $\Sigma_{xx}$. Note that the posterior covariance does not depend on the specific realization of $y$. 

The Gauss-Bayes point estimator for the price prediction, the conditional mean $\widehat {z}_{z\mid y}$ , minimizes the mean squared error of the estimate \cite{scharf1991statistical}. It turns out that for a specific observation $y$, the inverse of the conditional covariance is the Fisher Information matrix associated with estimating $z$ from $y$, and therefore $ \widehat \Sigma _{z\mid y}$ is the lower bound on the error covariance matrix for any unbiased estimator of $z$. 

The same set of equations arise in Kalman filtering. Kalman’s own view of this process is as a completely deterministic operation \cite{byrnes1994nonlinear}, and does not rely on assuming normality.
Although the point estimator $\widehat {z}_{z\mid y}$  is optimal in term of MSE, in practice there are numerical complications involved in this method: The matrix   $\Sigma _{yy}$ is typically not well conditioned, so the numerical calculation of $\Sigma _{yy}^{-1}$ cannot always be trusted. To overcome this problem, we propose a better conditioned estimator, which has a behavior close to Gauss-Bayes. 

\subsubsection{Principal Components and Estimation in Lower Dimension}

Principal component analysis (PCA) is a well-established mathematical procedure for dimensionality reduction of data and has wide applications across various fields. In this work, we consider its application in forecasting stock prices.

Consider the singular value decomposition (SVD) of $\Sigma _{xx}$:
\begin{equation}
    \Sigma _{xx} = VSV',
\end{equation}

where $S$ is a diagonal matrix of the same dimension as $x$ with non-negative diagonal elements in decreasing order, and $V$ is a unitary matrix ($VV^{T}=I_{N}$). The diagonal elements of $S$ are the eigenvalues of $\Sigma _{xx}$.

In general, the first few eigenvalues account for the bulk of the sum of all the eigenvalues. The ``large'' eigenvalues are called the principal eigenvalues. The corresponding eigenvectors are the called the principal components.

Let $L<N$ be such that the first $L$ eigenvalues in $S$ account for the bulk part (say 85\% or more) of the sum of  the eigenvalues. Let $V_{L}$ be the first $L$ column of unitary matrix $V$. Then the random vector $x$ is approximately equal to the linear combination of the first $L$ columns of $V$:
\begin{align}\label{alpha}
x\approx V_{L}\alpha,
\end{align}
where $\alpha$ is a random vector of length $L$. Because $L$ is a small number compared to $N$, equation (\ref{alpha}) suggests that a less ``noisy" subspace with a lower dimension than $N$ can represent most of the information. Projecting onto this principle subspace can resolve the ill-conditioned problem of $\Sigma _{yy}$. The idea is that instead of including all eigenvalues in representing $\Sigma_{xx}$, which vary greatly in magnitude, we use a subset which only includes the ``large'' ones, and therefore the range of eigenvalues is significantly reduced. The same concept is implemented in signal subspace filtering methods, which are based on the orthogonal decomposition of noisy speech observation space onto a signal subspace and a noise subspace \cite{hermus2006review}. Let $V_{M,L}$ be the first $M$ rows and first $L$ columns of $V$. We have
\begin{align}
  y=V_{M,L}\alpha + \text{Noise}.
\end{align}
 Mathematically resolving noisy observation vector $y$ onto the principle subspace can be written as a filtering operation in the form of 
\begin{align}
  w=Gy,
\end{align}
where $G$ is given by
\begin{align}
  G=(V_{M,L}^{'}V_{M,L})^{-1}V_{M,L}^{'}.
\end{align}
The vector $w$ is actually the coordinates of the orthogonal projection of $y$ onto the subspace equal to the range of $V_{M,L}$. We can also think of $w$ as an estimate of $\alpha$ based on least squares. Substituting $y$ by $w$ in  (\ref{cond}) leads to a better conditioned set of equations:
\begin{align}\label{condu}
 \widehat {z}_{z\mid w}&=\Sigma _{zw}\Sigma _{ww}^{-1}w \nonumber\\
  \widehat \Sigma _{z\mid w}&=\Sigma _{zz}-\Sigma _{zw}\Sigma _{ww} ^{-1}\Sigma _{wz},
\end{align}
because the condition number of $\Sigma_{ww}$ is much lower than that of $\Sigma_{yy}$,  as we will demostrate later. In (\ref{condu}) we have
\begin{align}
 \Sigma _{zw}= E\left [zw^{'} \right ]=\Sigma_{zy}G^{'},
\end{align}
and
\begin{align}
 \Sigma _{ww}= E\left [ww^{'} \right ]=G\Sigma_{yy}G^{'}.
\end{align}

If the posterior distribution of $z$ estimated based on (\ref{condu}) has a similar behavior to the distribution estimated by (\ref{cond}), it can be considered a good substitute for the Gauss-Bayes method. Our simulation demonstrates that this is indeed the case, which we will show in Section \ref{part2}.

\subsubsection{Unconditional Estimate}
Perhaps the simplest estimator for $z$ is just to use the (unconditional) mean and its associated covariance:
\begin{align}
\widehat{z}_\text{Uncon} &= E\left [z \right ] \nonumber\\
\widehat\Sigma _\text{Uncon} &=  \Sigma_{zz},
\end{align}
where $\widehat{z}_\text{Uncon}$ is the (unconditional) mean and $\widehat\Sigma _\text{Uncon}$ is the sample covariance of $z$.
Though this estimator ignores the historical prices in $y$, it is useful for comparison purposes, as will see in Section~\ref{C} below.

\subsection{Mean Squared Error}
To compare the performance of the methods discribed above, we evaluate the expected value of the squared error between the actual and estimated values. The mean squared error of an estimate $\hat{z}$ is given by:
\begin{align}\label{MSE}
MSE=&E\begin{bmatrix}
\left \| z-\hat{z} \right \|^{2}
\end{bmatrix}\nonumber\\ 
=&E\begin{bmatrix}
\left \| z \right \|^{2}
\end{bmatrix}
+E\begin{bmatrix}
\left \| \hat{z} \right \|^{2}
\end{bmatrix}-2E\begin{bmatrix}
\left \| z'\hat{z} \right \|
\end{bmatrix}.
\end{align}
The MSE can be expressed in terms of the covariance matrices in (\ref{sigmas}), by substituting the appropriate form of $\hat{z}$. In the unconditional case, when the unconditional mean is used as an estimate for future values, the point estimator is $\hat{z}=E\left [ z \right ]$. By substituting  $\hat{z}$ in (\ref{MSE}), the MSE will become

\begin{equation}
MSE_{Unc}=trace (\widehat\Sigma _\text{Unc}).
\end{equation}
In the Gauss-Bayes method, the point estimator is $\widehat {z}=\Sigma _{zy}\Sigma _{yy}^{-1}y$, which leads to
\begin{align}
MSE_{GB}=trace(\widehat\Sigma _\text{Unc})-trace(\Sigma_{zy}(\Sigma_{zy}\Sigma_{yy}^{-1})').
\end{align}
Finally in our reduced-dimension method using principle components, the point estimator can be formulated as $\widehat{z}=Cy$, where $C=\Sigma _{zw}\Sigma _{ww}^{-1}G$. The mean squared error then becomes:
\begin{align}
MSE_{RD}=trace(\widehat\Sigma _\text{Unc})+trace(C\Sigma_{yy}C')\nonumber\\
\mbox{}- 2trace(\Sigma_{zy}C').
\end{align}

Alternatively, the mean squared error of an estimator $\widehat{z}$ can be written in terms of the variance of the estimator plus its squared bias. The conditional MSE given $x$ is written as
\begin{align}\label{MSEY}
MSE_{\widehat{z}\mid z}=E\begin{bmatrix}
\left \| z-\hat{z}\right \| ^{2} \mid z
\end{bmatrix}= trace(\Sigma_{\widehat{z} \mid z}) \nonumber\\
+\left \| E\begin{bmatrix} \hat{z} \mid z \end{bmatrix} - z\right \| ^{2}.
\end{align}
The first term is called the variance, and the second term is the squared bias. The expected value of MSE over all observations is the actual (unconditional) MSE, which can be calculated by taking expectations on both sides of (\ref{MSEY}):
\begin{align}
MSE=E\begin{bmatrix}
E\begin{bmatrix}
\left \| z-\hat{z}\right \| ^{2} \mid z
\end{bmatrix} \end{bmatrix}= trace(E\left \| \Sigma_{\widehat{z} \mid z} \right \|) \nonumber\\
+ E\begin{bmatrix}
\left \| E\begin{bmatrix} \hat{z} \mid z \end{bmatrix}
- z\right \| ^{2}\end{bmatrix}.
\end{align}

It turns out that Gauss-Bayes estimator is unbiased, which means that the second term is 0, while the unconditional estimator and the proposed reduced-dimension methods are biased estimators as we will show next.

We can calculate the bias as follows. In Gauss-Bayes, by writing the estimator as $\widehat{z}=Dy$ where $D=\Sigma _{zy}\Sigma _{yy}^{-1}$, the bias is:
\begin{align}
Bias_\text{GB}=&E\begin{bmatrix}
\left \| E\begin{bmatrix} E[z \mid y] \mid z \end{bmatrix}
- z\right \| ^{2}\end{bmatrix}\nonumber \\
=&E\begin{bmatrix}
\left \| z
- z\right \| ^{2}\end{bmatrix} = 0.
\end{align}
As it was mentioned above, the Gauss-Bayes estimator is unbiased and in fact the covariance matrix in this method provides the lower limit on the MSE for all unbiased estimators. 
In unconditional case:
\begin{align}
Bias_\text{Unc}=& E\begin{bmatrix}
\left \| E\begin{bmatrix} E[z] \mid z\end{bmatrix} - z\right \| ^{2}\end{bmatrix}
\nonumber\\
=& E\begin{bmatrix}
\left \| E\begin{bmatrix} z \end{bmatrix}
- z\right \| ^{2}
\end{bmatrix}
=trace( \Sigma_{zz} ).
\end{align}

In this case the bias is equal to the MSE value, which means the variance of the estimator is zero. This is because in this case, we have the same estimator for different observations.
In the reduced-dimension method, the bias in terms of $C$ is calculated as:
\begin{align}
Bias_{RD}=&E\begin{bmatrix}
\left \|C E\begin{bmatrix} y \mid z \end{bmatrix}
- z\right \| ^{2}\end{bmatrix} 
\nonumber\\
=&trace\begin{bmatrix} (I-CR)\Sigma_{zz}(I-CR)^{'}\end{bmatrix}
\end{align}
where $R=\Sigma _{yz}\Sigma _{zz}^{-1}$.

\subsection{Numerical Illustration}\label{C}
\begin{figure}
 \begin{center}
\includegraphics[width=8.5cm]{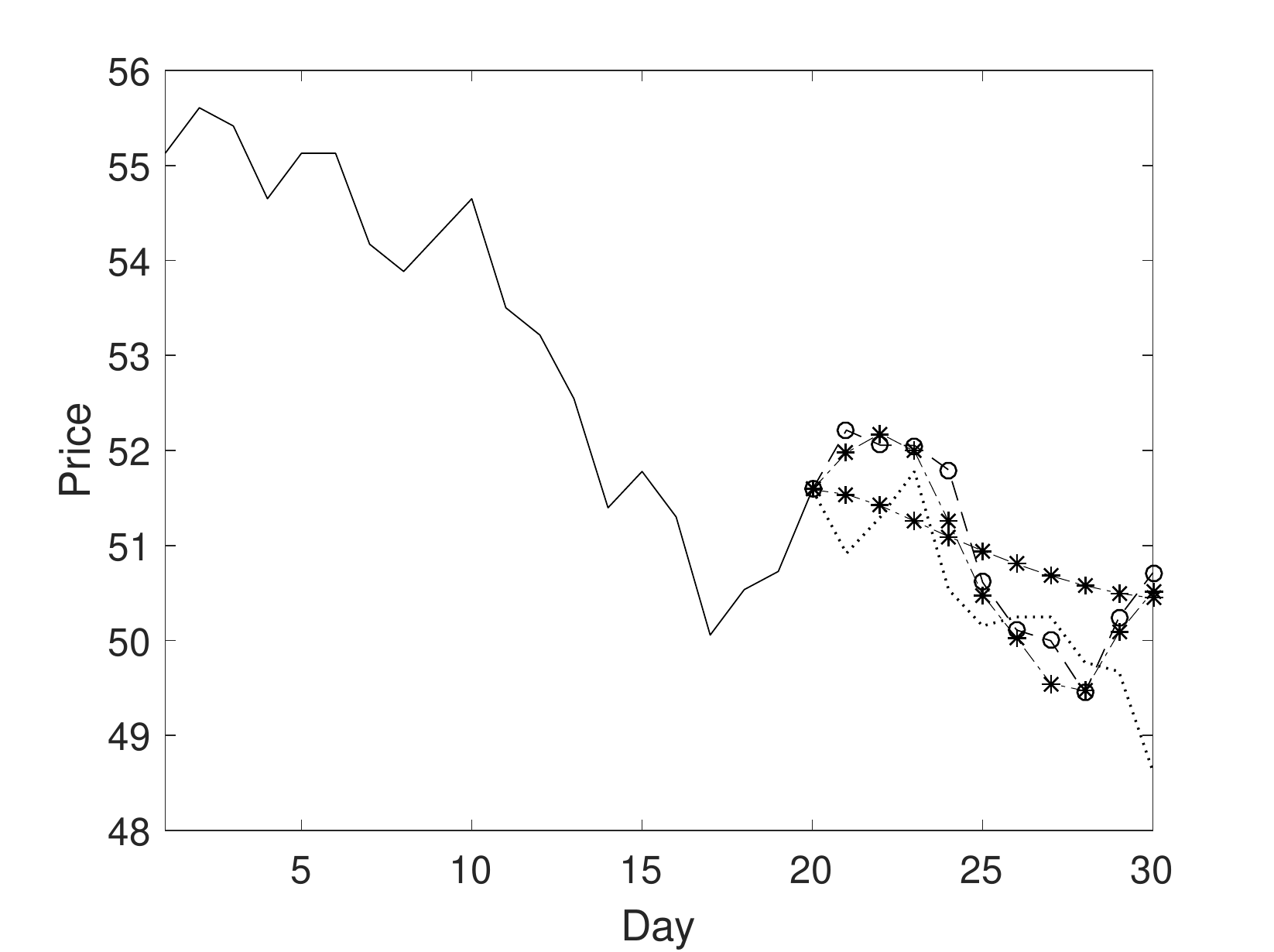}
\caption{Predicting price for $M+1$ to $N$ days, Actual Price - GB --o RD --* }
\label{fig:Price}
\end{center}
\end{figure}

To illustrate the discussion thus far, we implement the three different techniques of estimation for the next 10 days (day $M+1$ to $N$ where $N-M=10$). Figure ~\ref{fig:Price} shows an example of our stock predictions. Assume that we are given the price values for the past 20 days $M=20$, and we want to use those values to predict the future prices over the next 10 business days, from day $M+1$ to day $N$, $N=30$. In our reduced-dimension technique, we can get a relatively smooth plot of the predicted value for a relatively small $L$, to a plot almost the same as Gauss-Bayes, for larger values of $L$.

As mentioned before, the mean squared error (MSE) is a common and appropriate measure of performance. Equations for calculating MSE in each case were introduced in previous sections. For a given (estimate of) covariance matrix, we implement our reduced-dimension technique for different $M$s, and for different numbers of principal eigenvalues, $L$. The graphs in this section are only for the purpose of illustration. The performance of the methods will be evaluated more comprehensively in the next section. The data used in this section were normalized and centered, as described in detail later.

\begin{figure}
 \begin{center}
\includegraphics[width=8.5cm]{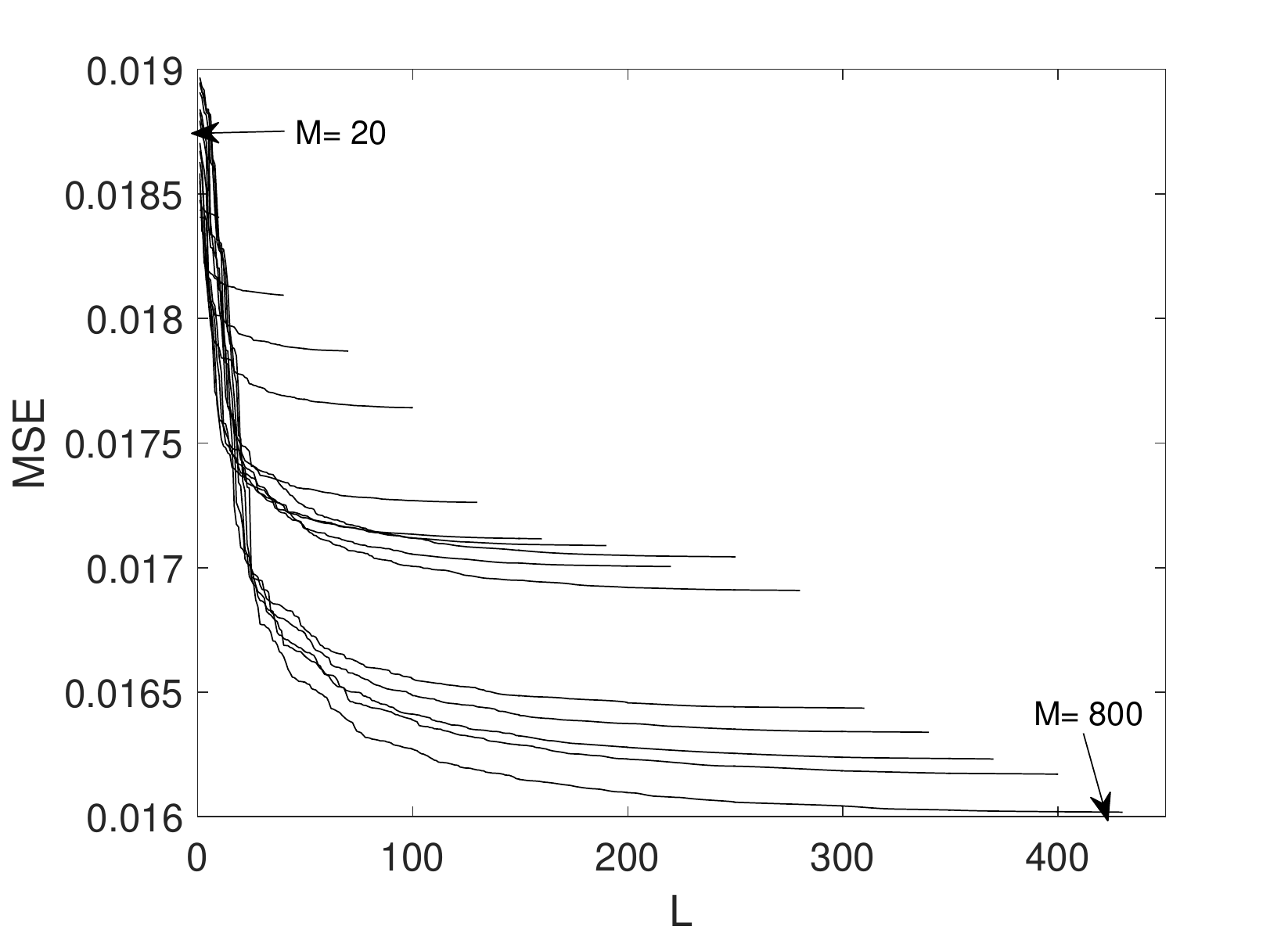}
\caption{MSE versus $L$ in the normalized domain for different $M$s}
\label{fig:MSEUvsL}
\end{center}
\end{figure}

To illustrate the performance of the methods, we calculate the sum of MSE values over all days of estimation. We use General Electric company price data over a period of 5000 days to calculate the MSE values illustrated in this section.
Figure~\ref{fig:MSEUvsL} shows the values of MSE over all days of estimation versus value of $L$, for 27 different lengths of observation vector $M$, from 20 to 800. It turns out that MSE value is not that sensitive to the value of $L$ for sufficiently large $L$. As we can see, initially, the MSE values fall quickly for small values of $L$, but then remain relatively constant, so if we have a particular constraint on condition number, we do not lose that much in terms of MSE by choosing a reduced-dimension subspace, which leads to a better conditioned problem.

\begin{figure}
 \begin{center}
\includegraphics[width=8.5cm]{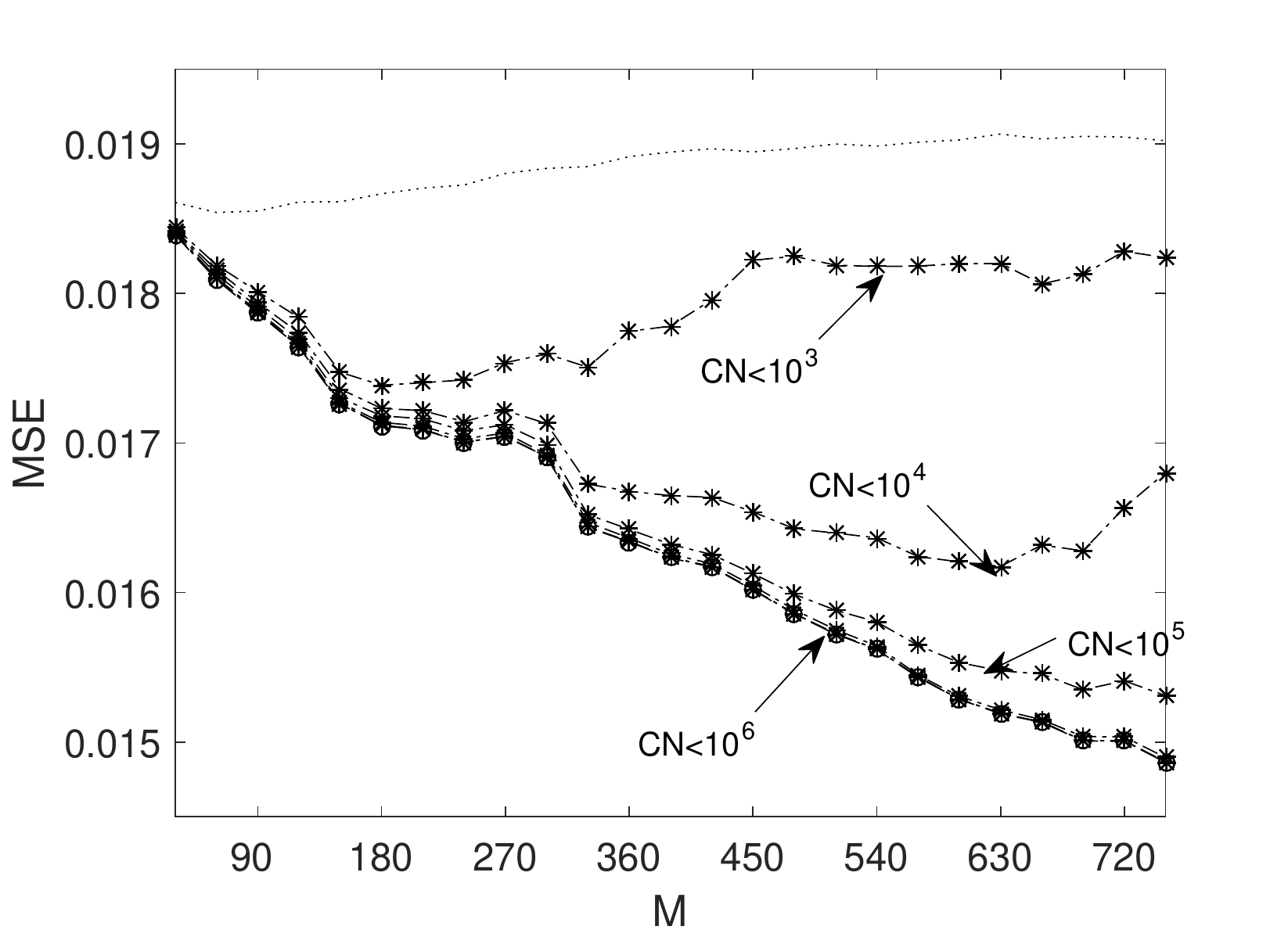}
\caption{Best MSE subject to different upper limit on condition number, Unc ... GB --o RD --* }
\label{fig:MSEU}
\end{center}
\end{figure}

Figure~\ref{fig:MSEU} shows the sum of MSE over all days for our reduced-dimension method, subject to an upper limit on condition number of $\Sigma_{ww}$. 
There is a trade-off between $M$, length of observation vector, and value of MSE. In general, by increasing $M$, more information is available in each observation, resulting in better performance of the prediction. For each length of $M$, the values for MSE are captured based on different constraints of the condition number of $\Sigma_{ww}$. The top plot in Figure~\ref{fig:MSEU} corresponds to the MSE values corresponding to the Unconditional method. The plot on the bottom corresponds to Gauess-Bayes' MSE values, which indicates the optimal performance. The other 4 plots correspond to our reduced-dimension method, subject to 4 different upper limits on $\Sigma_{ww}$ condition number, from $10^{3}$ to $10^{6}$. We should point out that the condition number of the $\Sigma_{yy}$ matrices used in these plots were sufficiently small that the calculations for Gauss-Bayes were considered reliable. This is not the case in general for practical situations.

The performance of the reduced-dimension estimation method is close to Gauss-Bayes in terms of MSE up to a certain point for all different limits on condition number. After $M=200$, or in other words after roughly 7 months, the values from reduced-dimension methods start deviating from Gauss-Bayes in some cases. The first line on top is the best performance of reduced-dimension method subject to condition number of $\Sigma_{ww}$ being less than $10^{3}$ which is about 1000 times better than the condition number of $\Sigma_{yy}$. However, in this case after approximately $M=360$, the performance of reduced-dimension estimator starts deteriorating. This illustrates that it is important to choose an appropriate value of $M$.
A possible explanation for this observation is that when we fix some constraint on condition number, we are actually limiting the value of $L$, and by increasing $M$, after a certain point, we mostly increase the noise, and the MSE value gets worse. 

\begin{figure}
 \begin{center}
\includegraphics[width=8.5cm]{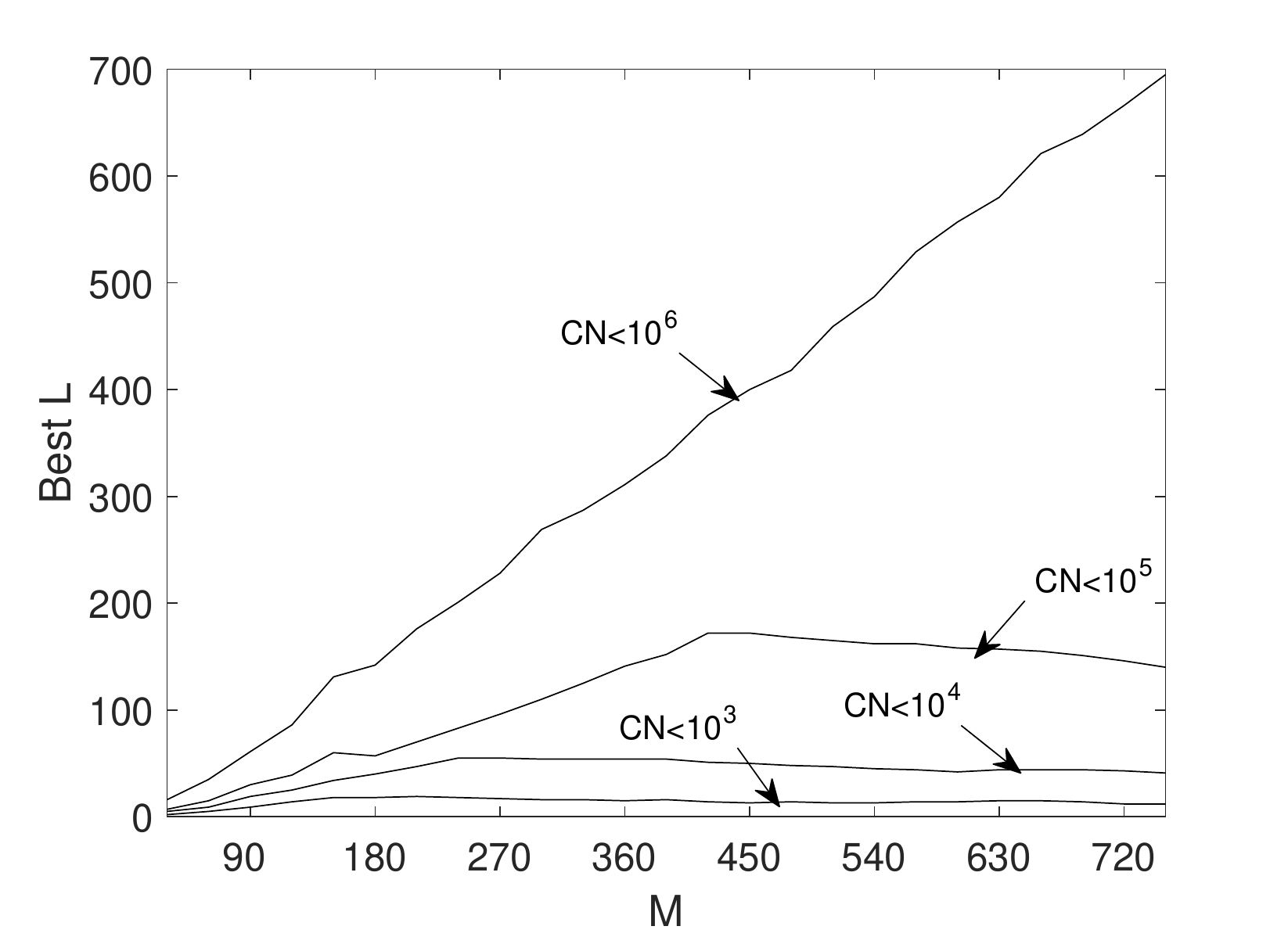}
\caption{Best $L$ subject to different limits on condition number}
\label{fig:BestL}
\end{center}
\end{figure}

Recall that $L$ represents the number of eigenvalues required from the diagonal matrix $S$ to represent the bulk part of the information carried in $x$.
Figure~\ref{fig:BestL} shows the value of $L$ corresponding to best MSE for different $M$s, subject to different limits on condition number. As the upper limit on condition number increases, the value of MSE improves as $M$ increases, and we need a bigger subspace, bigger $L$, to extract the information. However, as the bottom three plots in Figure~\ref{fig:BestL} show, the value for best $L$ is almost constant after a certain point, which is consistent with Figure~\ref{fig:MSEUvsL}.

\section{Empirical Methodology and Results}\label{part2}
In this section we describe how we estimate the covariance matrix based on normalized data set, and we evaluate the performance of our method using empirical data. 

\subsection{General Setting}
Suppose that we have $K$ samples of vector data, each of length $N$, where $N < K$. Call these row vectors $x_{1},x_{2},...,x_{K}$, where each $x_{i}\in \mathbb{R}^{N} (i=1,...,K)$ is a row vector of length $N$: 
\begin{equation}
x_{i}=\begin{bmatrix}
x_{i1} & x_{i2} & \cdots & x_{iN}\\
\end{bmatrix}.
\end{equation}
We assume that the vectors $x_{1},x_{2},\ldots,x_{K}$ are drawn from the same underlying distribution. We can stack these vectors together as rows of a $K\times N$ matrix: 

\[
X=\left[ \begin{array}{cccc}
                x_{11} & x_{12} & \cdots & x_{1N} \\
               x_{21} & x_{22} & \cdots & x_{2N} \\
            \cdots & \cdots & \cdots & \cdots \\ 
                 x_{K1} & x_{K2} & \cdots & x_{KN}
              \end{array}
\right].
\]

Let $M\leq N$ and suppose that we are given a vector $y\in \mathbb{R}^{M}$ representing the first $M$ data points of a vector we believe is drawn from the same distribution as $x_{1},x_{2},\ldots,x_{K}$. Again, these $M$ data points represent the end-of-day prices of a company stock over the past $M$ consecutive trading days. Let $z$ be the price of the next $N-M$ days in the future. We wish to estimate $z$ from $y$.

Since the vector $x_{i}$ is a multivariate random vector that can be partitioned in the form
\begin{equation}
x_{i}=\begin{bmatrix}
y_{i} &
z_{i}
\end{bmatrix},
\end{equation}
where $y_{i}$ has length $M$ and $z_{i}$ has length $N-M$, accordingly the data matrix $X$ can be divided into two sub-matrices $Y$ and $Z$ as follow: 
\[
\left[ \begin{array}{c}
             X\\
              \end{array}
\right]=
\left[ \begin{array}{c c}
            Y & Z\\
              \end{array}
\right].
\]
We can think of $Y$ as a data matrix consisting of samples of historical data, and $Z$ as a data matrix consisting of the corresponding future values of prices.
\subsection{Normalizing and Centering the Data}
In the case of stock-price data, the vectors $x_{1},x_{2},\ldots,x_{K}$ might come from prices spanning several years or more. If so, the basic assumption that they are drawn from the same distribution may not hold because the value of a US dollar has changed over time, as a result of inflation.
To overcome this issue, a scaling approach should be used to meaningfully normalize the prices. One such approach is presented here. Suppose that $t_{i}=[t_{i1},t_{i2},\ldots,t_{iN}]'$ is a vector of ``raw" (unprocessed) stock prices over $N$ consecutive trading days. Suppose that $Q\leq N$ is also given. Then we apply the following normalization to obtain $x_{i}$: 
\begin{align}
    x_{i}=\frac{t_{i}}{t_{i}(Q)}.
\end{align}
This normalization has the interpretation that the $x_{i}$ vector contains stock prices as a fraction of the value on the $Q$th day, and is meaningful if we believe that the pattern of such fractions over the days $1,\ldots ,N$ are drawn from the same distribution. Note that $x_{i}(Q)= 1$. 

For the purpose of applying our method based on PCA, we assume that the vectors $x_{1},x_{2},\ldots,x_{K}$ are drawn from the same underlying distribution and that the mean, $\bar{x}$, is equal to zero. However because $x_{i}$ represents price values, in general the mean is not zero. The mean $\bar{x}$ can be estimated by averaging the vector $x_{i}\in \mathbb{R}^{N} (i=1,...,K)$,
\begin{align}
\bar{x} &= \frac{1}{K} \sum_{i=1}^{K} x_{i},
\end{align}
and then this average vector is deducted from each $x_{i}$ to center the data.

\section{Experiments}

The daily historical price data from 1966 to 2015 for General Electric Company (GE) was downloaded from finance.yahoo.com. This data set is transformed into a Hankel matrix (defined below) and then centered and normalized to construct the data matrices, as described earlier. The estimation results for stock price estimation of other companies from different industries were also included in this study which includes: Exxon Mobil Corporation (XOM), Wal-Mart Stores, Inc. (WMT), Intel Corporation (INTC), and Caterpillar Inc. (CAT) from farm and construction machinery. SPDR S\&P 500 ETF (SPY), which is a fund, following the S\&P500 index, was also included. The number of observations for these companies might be fewer than the number of observations for GE because less historical price data is available. In this paper we focus on short-term prediction, meaning just a few days. We compare the estimation methods based on their out-of-sample performance.

\subsection{Constructing Data Matrix}

The daily stock price data is transformed into a matrix with $K$ rows, samples of vector data, each of length $N$. We get that by stacking $K$ rows ($K$ samples), each one time shifted from the previous one, all in one big matrix, called the Hankel matrix.

More precisely, the Hankel matrix for this problem is constructed in the following format:
\[
\begin{bmatrix}
x_{1}\\ 
x_{2}\\ 
\vdots\\
x_{K}\\ 
\end{bmatrix}= 
\begin{bmatrix}
  P (1) & P (2) & \cdots &  P (N)\\ 
 P (2)& P (3) & \cdots & P (N+1) \\ 
 \cdots & \cdots & \cdots &  \cdots \\ 
 P (K) & P (K+1) & \cdots &  P (K+N-1)
\end{bmatrix},
\]

where $P(i)$ represents the price for day $i$. This is our matrix of data, $X$, before normalization and centering.

As explained earlier, we first normalize each row by $x(Q)$ and then subtract the average vector $\bar{x}$ from each row. After running the simulation, to make use of the predicted values, we should add back the average vector $\bar{x}_{N-M}$ (last $N-M$ components of $\bar{x}$) from days $M+1$ through $N$ and also multiply the result by $x(Q)$ to get back to actual stock prices.
We tested different values for $Q$ in terms of MSE and estimation variance. For the purpose of this study, we chose $Q=M$ because it shows the best results in this setting. 
Note that $x_{i}(M)= 1$. This column is removed from the data matrix because it does not provide any information. From now on matrix $X$ represents normalized and centered price data. Then the sample covariance matrix is calculated as $\Sigma_{xx}=X^{T}X/(N-1)$.

We obtained end-of-day stock prices for General Electric company for about 12500 consecutive days. We then converted this time series into Hankel matrices with different lengths as described above. We divided the data vectors into two parts: the first was used to estimate $\Sigma_{xx}$ and the second was used for performance evaluation. We included 12300 sampled in the data matrix, and 2200 samples are used to evaluate the out-of-sample performance of the methods. We vary $M$ from 20 to 440, with a 30 day interval, to investigate the effect of length of observation vector on the results, which means 15 sets of data were evaluated in this study. 

\begin{figure}
 \begin{center}
\includegraphics[width=8.5cm]{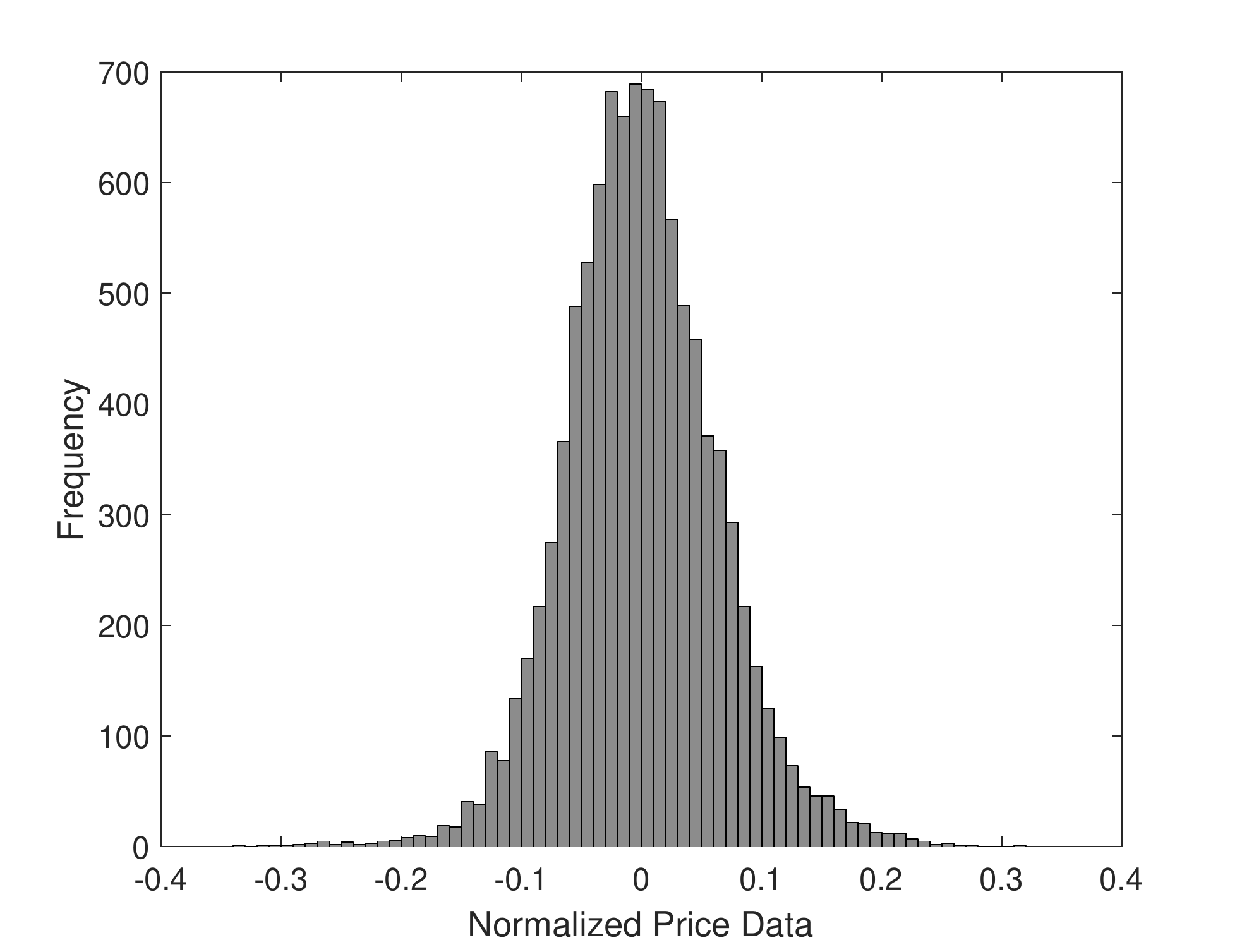}
\caption{Histogram graph for normalized data}
\label{fig:histogram}
\end{center}
\end{figure}

The histogram of normalized data is graphed as a representation of the distribution of data. Figure~\ref{fig:histogram} represent the first predictor (first column) in matrix $X$, and the curve resembles a bell shape.

\subsection{MSE Performance}

For each of the data sets constructed above, we implement the three different techniques of estimation described earlier for the next 10 days (day $M+1$ to $N$). 

The general goal, as mentioned above, is an estimation technique that has a similar behavior as Gauss-Bayes but does not have the associated calculation difficulties resulting from ill-conditioning. As mentioned before, mean squared error (MSE) is a common and appropriate measure of performance. We take the average of MSE values over 2200 samples to evaluate the performance of the methods. We implement our reduced-dimension technique for different $M$s, and for different numbers of principal eigenvalues, $L$. 

\begin{figure}
 \begin{center}
\includegraphics[width=8.5cm]{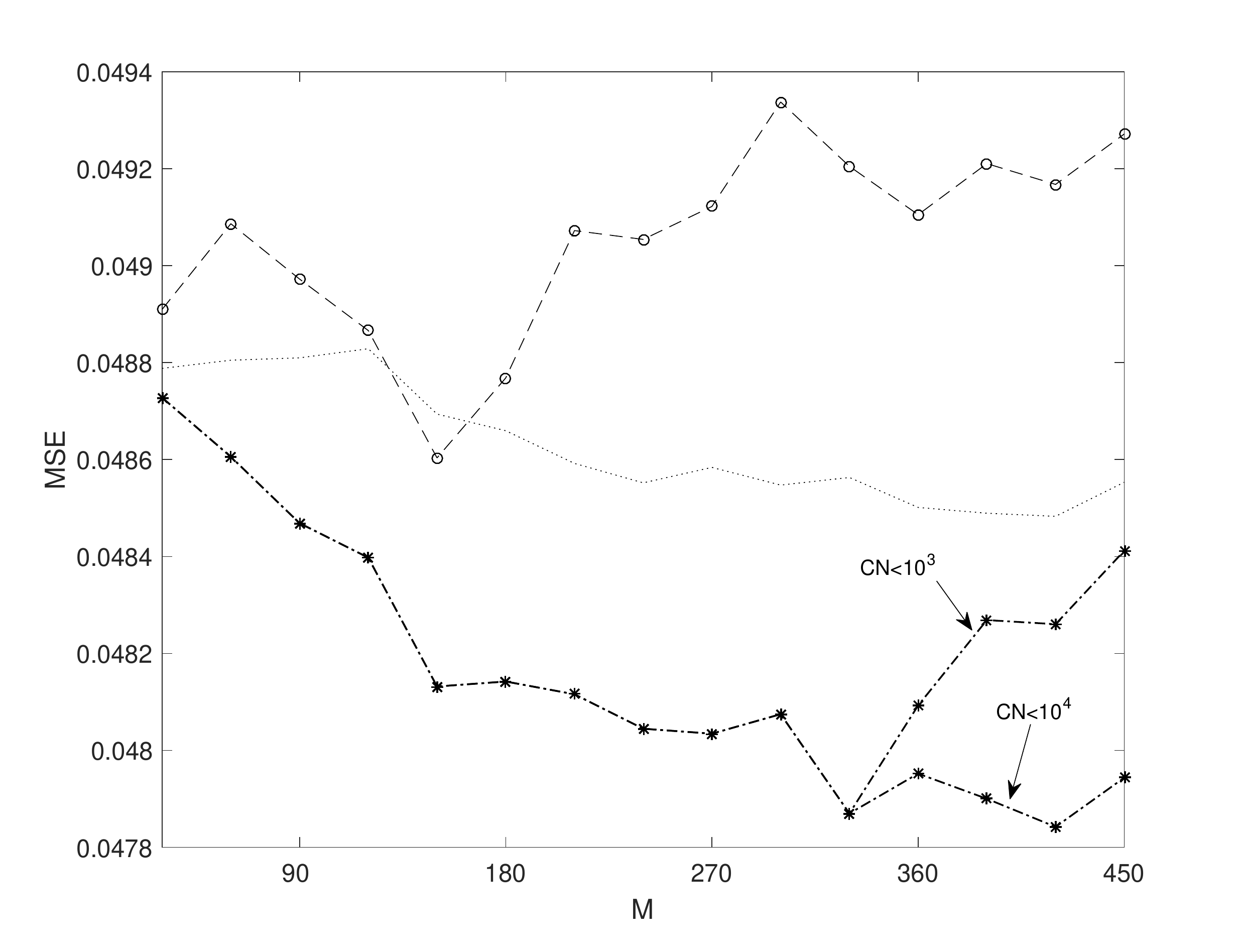}
\caption{Best MSE subject to different upper limit on condition number of $\Sigma_{ww}$, Unc ... GB --o RD --*}
\label{fig:MSEU2200}
\end{center}
\end{figure}

We are interested in the sum of MSE values over all days of estimation as shown in Figure~\ref{fig:MSEU2200}. The figure illustrates the sum of MSE over all days, subject to an upper limit on condition number of $\Sigma_{ww}$. 
As illustrated before, there is a trade-off between $M$, length of observation vector, and value of MSE. In general, by increasing $M$, more information is available in each observation, resulting in better performance of the estimation. The almost flat plot in the middle of Figure~\ref{fig:MSEU2200} corresponds to the MSE values for the Unconditional method, which is based on the empirical mean of our normalized and centered data set. The plot on the top corresponds to Gauss-Bayes' MSE values. When it comes to out-of-sample performance, the numerical complications overpower the estimation accuracy of Gauss-Bayes, causing the MSE plot for this method to lie even higher than the MSE plot for the Unconditional estimator. 
Finally, the plots at the bottom correspond to our reduced-dimension method, subject to two different upper limits on the condition number of $\Sigma_{ww}$, $10^{3}$ and $10^{4}$. In this case the MSE values do not improve by increasing the upper limit on condition number of $\Sigma_{ww}$, so higher condition number limits are not illustrated in Figure~\ref{fig:MSEU2200}. Even the two plots illustrated here are almost the same except for after $M$ exceeds roughly 350 days.

As discussed before, when we fix some constraint on condition number, we are actually limiting the value of $L$, and by increasing $M$, after a certain point, we mostly increase the noise, and the MSE value gets worse. That is why in the plot for best performance of reduced-dimension method subject to condition number of $\Sigma_{ww}$ being less than $10^{3}$, in Figure~\ref{fig:MSEU2200}, the value of MSE starts increasing after a certain point.

\begin{figure}
 \begin{center}
\includegraphics[width=8.5cm]{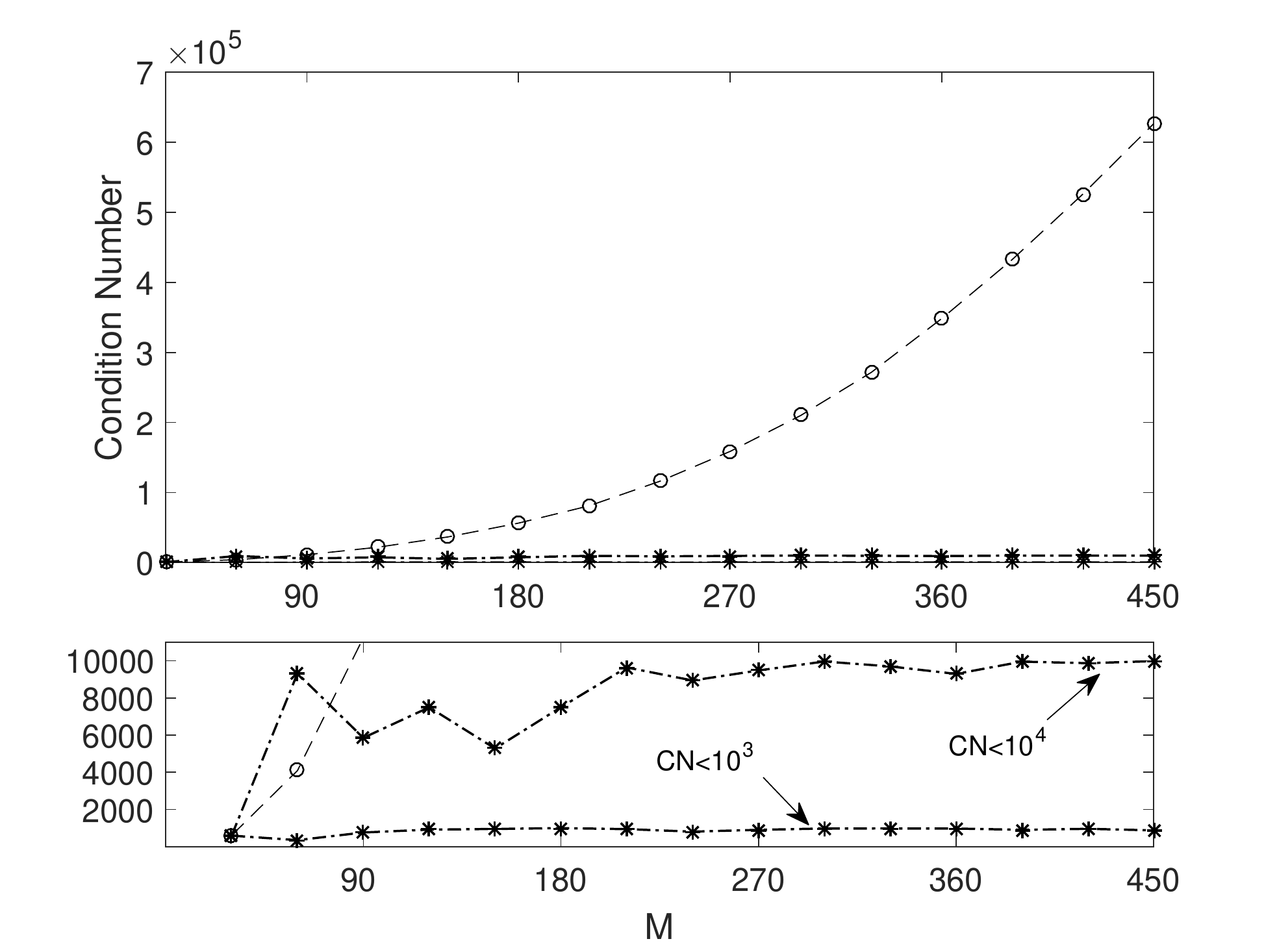}
\caption{Condition number for different $M$s, Plot on the bottom magnifies the lower section of the plot on top, GB --o RD --*}
\label{fig:Cond}
\end{center}
\end{figure}

Figure~\ref{fig:Cond} shows the values of condition number for Gauss-Bayes and the reduced-dimension method. 
As shown in the graph, the reduced-dimension method improves the condition number of the problem by orders of magnitude, resulting in a better performance. In both cases, as $M$ increases, the condition number increases. This is consistent with the fact that for a bigger $M$ we need a bigger subspace to extract the information effectively, which results in a bigger condition number. Table~\ref{tab:MSEtable} shows the MSE values. The reduced-dimension method shows better performance than the other two methods.

\begin{table}
 \caption{MSE for different companies (M=330)}
\label{table1}
\centering
\begin{tabular}{c||c c  c}

MSE & Unc & GB  & RD
\\    \hline
GE & 0.002358378
 & 0.002421157
 & 0.00229152 
\\    
INTC & 0.00171524
 & 0.001767552
  & 0.001696432
\\ 
XOM & 0.000970769
 & 0.000964532
  & 0.000920936
\\
CAT & 0.002761635
 & 0.002840376
  & 0.00269465
\\
WMT & 0.00064775
 & 0.000687175
  & 0.000629935
\\
SPY & 0.001798499
 & 0.001904198
  & 0.00170225
\\
\end{tabular}
 \label{tab:MSEtable}
\end{table}

\begin{figure}
 \begin{center}
\includegraphics[width=8.5cm]{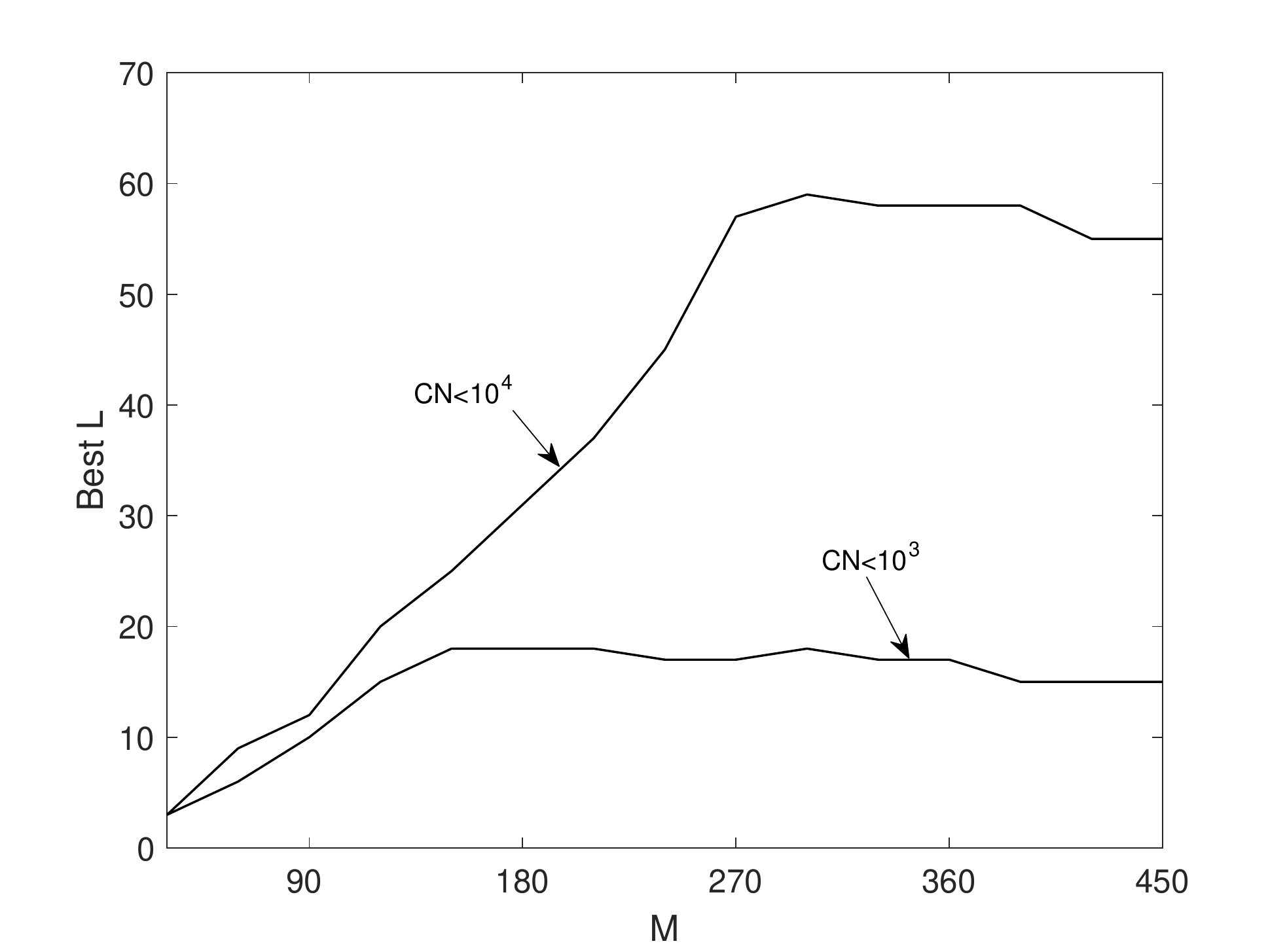}
\caption{Best $L$ subject to different limits on condition number}
\label{fig:BestL2200}
\end{center}
\end{figure}

Figure~\ref{fig:BestL2200} investigates the dimension of the target subspace by plotting the value of $L$ corresponding to best MSE for different $M$s, subject to different limits on condition number (the same case as in Figure~\ref{fig:MSEU2200}). Again, $L$ represents the number of eigenvalues required from the diagonal matrix $S$ to represent the bulk part of the information.

As the upper limit on condition number increases, the value of MSE improves as $M$ increases, and we need a bigger subspace, bigger $L$, to extract the information.

\subsection{Directional Change Statistic}
\begin{figure}
  \centering
  \includegraphics[width=8.5cm]{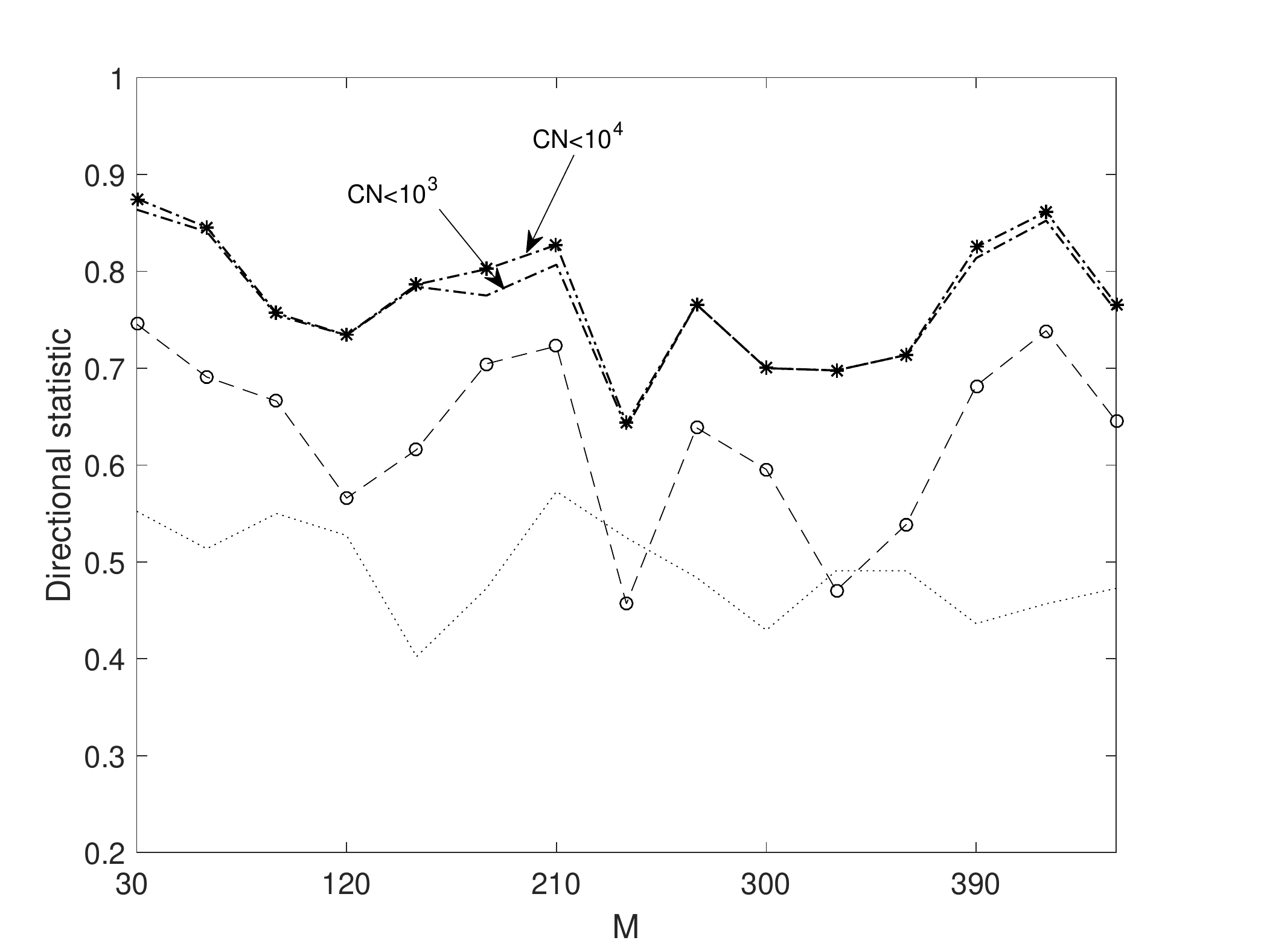}
  \caption{Best Directional Statistics subject to different upper limit on condition number of $\Sigma_{ww}$, Unc ... GB --o RD --*}
  \label{fig:DirStat}
\end{figure}

Another evaluation metric of interest is called the directional statistic which measures the matching of the actual and predicted values in terms of directional change. To be precise, let $\hat{z}_{i}=\begin{bmatrix}
\hat{z}_{i(M+1)} & \hat{z}_{i(M+2)} & \cdots & \hat{z}_{iN}\\
\end{bmatrix}$ be the price prediction for the next $M+1$ to $N$ days. We evaluate the direction of the prediction compared to $z_{0}$ or today's price. For $j= M+1, M+2,...,N$ we have:
 \begin{equation}
    b_{ij}=
    \begin{cases}
      1, & \text{if}\ (z_{ij}-z_{0})(\hat{z}_{ij}-z_{0})>0 \\
      0, & \text{otherwise}
    \end{cases}
  \end{equation}
and then $D_{j}$, the direction statistic for day $j$, averaged over $K$ samples, is equal to
 \begin{align}
D_{j} &= \frac{1}{K} \sum_{i=1}^{K} b_{ij},
\end{align}

which is a number between 0 and 1 (the higher the better). Figure~\ref{fig:DirStat} shows the average of directional statistic over 10 days of estimation using the same $K=2200$ samples. As the plot indicates, the reduced-dimension method is superior in terms of directional change statistic. It is interesting to note that the directional statistic does not significantly depend on $M$.

Table~\ref{tab:DirStat} represent the directional statistic for different companies and SPY for $M=390$ for $\Sigma_{ww}$ condition number limited to $10^{4}$. The reduced-dimension method is superior to the other two methods in terms of directional change estimation. It is important to notice that the values represented in Table~\ref{tab:DirStat} are associated with a specific $M$ for all different cases. In practice, it is recommended to choose the appropriate $M$ for each stock to get the best results. For example, for Intel, for $M=300$, the directional statistic is equal to 0.834 for reduced-dimension method, 0.66 for Gauss-Bayes, and 0.4 for Unconditional estimation, which are better than the values in the table.

\begin{table}
 \caption{Directional Statistics for M=390}
\label{table1}
\centering
\begin{tabular}{c||c c  c}

Directional Statistic & Unc & GB  & RD
\\    \hline
GE & 0.436363636
 & 0.681818182
 & 0.825
\\    
INTC & 0.525
 & 0.540909091
  & 0.715909091
  \\
XOM & 0.452272727
& 0.681818182
& 0.856818182
  \\
CAT & 0.561363636
& 0.468181818
& 0.786363636
  \\
WMT & 0.388636364
& 0.429545455
& 0.795454545
\\ 
SPY & 0.431818182
 & 0.484090909
  & 0.731818182
\\
\end{tabular}
 \label{tab:DirStat}
\end{table}

\subsection{Volatility}
Another important parameter that we estimate is the volatility of the prediction, measured in terms of its standard deviation. 
The squared root of diagonal elements of the estimated covariance, $\widehat \Sigma_{zz}$, are the estimated standard deviations for individual days of estimation. The estimate of the covariance in each method is
\begin{align}\label{stdformulas}
 \widehat \Sigma _{GB}&=\widehat \Sigma _{z\mid y}=\Sigma _{zz}-\Sigma _{zy}\Sigma _{yy}^{-1}\Sigma _{yz},
\nonumber\\
 \widehat \Sigma _{RD}&=\widehat \Sigma _{z\mid w}=\Sigma _{zz}-\Sigma _{zw}\Sigma _{ww} ^{-1}\Sigma _{wz},
 \nonumber\\
\widehat\Sigma _\text{Unc} &=\Sigma_{zz},
\end{align}

\begin{figure}
  \centering
  \includegraphics[width=8.5cm]{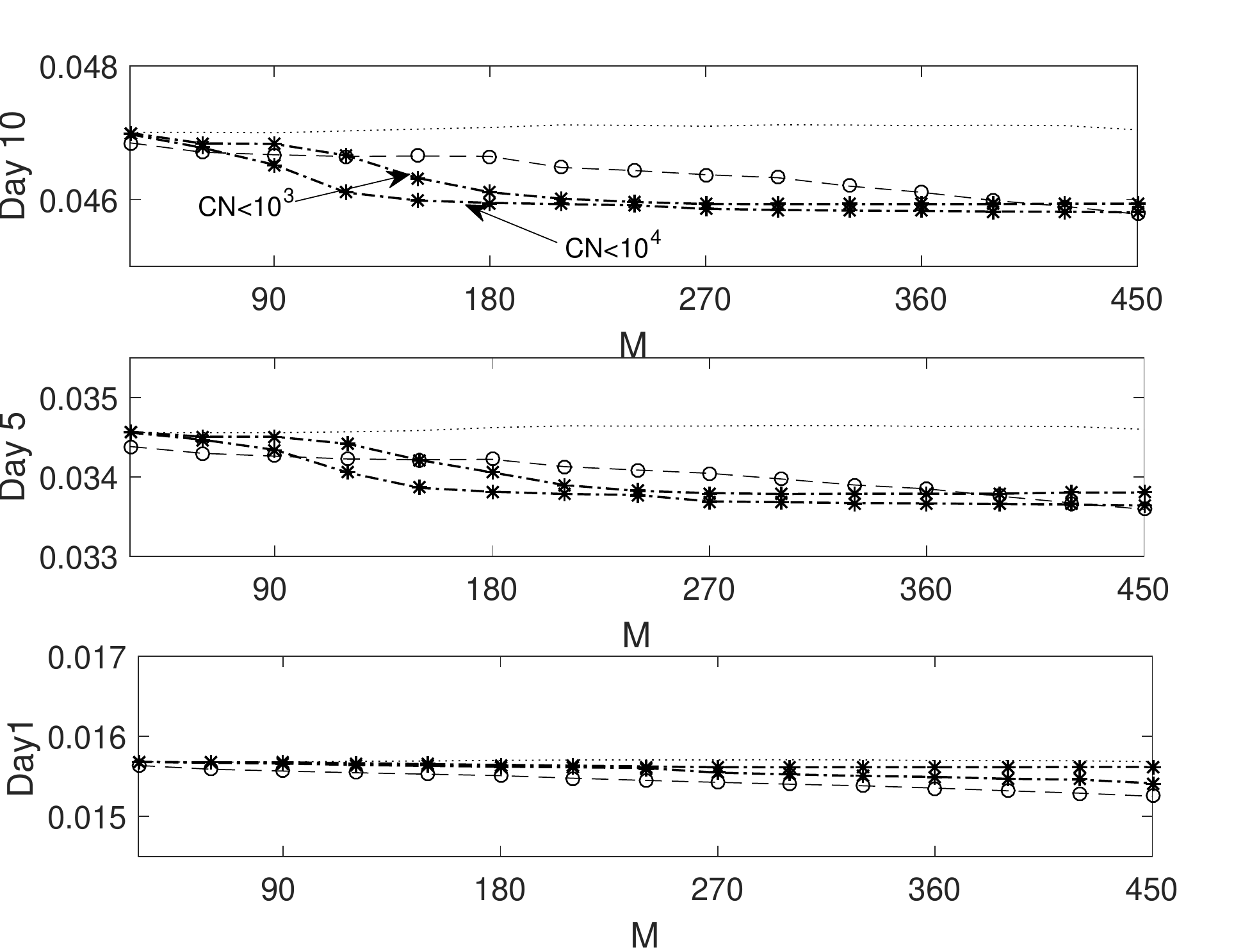}
  \caption{Standard deviation of individual days of estimation, Unc ... GB --o RD --*}
  \label{fig:Stdall}
\end{figure}

In general the standard deviation values increase moving from day 1 to day 10 of prediction, since less uncertainty is involved in the estimation of stock prices of days closer to the current day. Figure~\ref{fig:Stdall} illustrates an example corresponding to the case where the upper bounds on condition number in our reduced-dimension method are $10^{3}$ and $10^{4}$. The standard deviation for individual days of estimation are plotted versus $M$, the length observation vector, for Gauss-Bayes and the reduced-dimension method for days 1, 5, and 10 of estimation. The unconditional estimation is also included for comparison. The standard deviation values decrease as $M$ increases because more information is provided in each observation. For the first day of estimation, all three different lines are very close, but as we move up, the unconditional estimation deviates from other two methods especially as $M$ increases. As illustrated in Figure~\ref{fig:Stdall}, the standard deviation values in our reduced-dimension method are very close to those of Gauss-Bayes values, but after a certain point for days further away from current day, the values in the reduced-dimension method decrease faster and then stay the same, which means we can get the same values as Gauss-Bayes with not only a smaller condition number, but also a smaller $M$.

\section{Conclusion}
In this paper, we described a general method for prediction using covariance information. We illustrated our method on daily stock price values for General Electric and 5 other companies. The daily historical price data was transformed into Hankel matrices of 15 different lengths to investigate the impact of length of observation on estimation power. We describe our method for normalizing and centering the empirical. The multivariate conditional mean (Gauss-Bayes) is known to minimize the mean squared error and therefore is used as a suitable unbiased estimator of future values; however, the numerical results from this method cannot be trusted because the covariance matrix is not well conditioned. We proposed a filtering operation using principle component analysis to overcome this issue. Resolving the observed data set onto a principle subspace reduces the dimensionality of the problem and improves the condition number of the problem by orders of magnitude. The proposed method shows better out-of-sample performance compared to the multivariate conditional mean in terms of mean squared error and directional change statistic and therefore is considered a good substitute for Gauss-Bayes.

The proposed method is easily implemented and can be modified to include multiple predictors. The significance of the proposed approach will be even more apparent in estimating the future price values using multiple predictors because in that case, where observation vectors are mostly longer than the one in this study, it becomes almost impossible to rely on Gauss-Bayes due to the severe ill-conditioning of the covariance matrix.

\section*{References}
\bibliography{mybibfile}

\end{document}